# Electrically reconfigurable dipolar polaritons with highly tunable nonlinearity in a homobilayer microcavity

Baixu Xiang[1,*], Yubin Wang[1,*], Guihan Wen[1,*], Yitong Li[1], Hao Wen[1], Zengde She[2], Haiyun Liu[3], Kenji Watanabe[4], Takashi Taniguchi[5], Timothy C. H. Liew[6], Zhiyuan Sun[1,7], Qihua Xiong[1,3,7,†]

[1] State Key Laboratory of Low-Dimensional Quantum Physics and Department of Physics, Tsinghua University, Beijing, China.

[2] Department of Physics, University of Washington, Seattle, USA.

[3] Beijing Academy of Quantum Information Sciences, Beijing, China.

[4] Research Center for Electronic and Optical Materials, National Institute for Materials Science, Tsukuba, Japan.

[5] Research Center for Materials Nano architectonics, National Institute for Materials Science, Tsukuba, Japan.

[6] Division of Physics and Applied Physics, School of Physical and Mathematical Sciences, Nanyang Technological University, Singapore.

[7] Frontier Science Center for Quantum Information, Beijing, China.

* These authors contributed equally: Baixu Xiang, Yubin Wang, Guihan Wen.

To whom correspondence should be addressed. Email: qihua_xiong@tsinghua.edu.cn.

## Abstract

Active control over optical nonlinearity in solid-state systems is central to unlocking exotic many-body phenomena and photonic devices. While exciton-polaritons in transition metal dichalcogenides (TMDs) offer a promising platform, their practical utility is impeded by fixed interactions and a trade-off between nonlinearity and oscillator strength. Here, we report electrically tunable dipolar polaritons in a dual-gated bilayer $MoS_2$ microcavity, demonstrating *in situ* reshaping of the dispersion and modulation of the light-matter coupling strength *via* the quantum-confined Stark effect. Crucially, this electrical control yields a seven-fold enhancement of the polariton-polariton interaction strength. This enhancement arises from the combined tuning of the effective exciton-exciton interaction and the excitonic Hopfield coefficient. In addition, electrostatic doping provides an independent knob to continuously drive a strong-to-weak coupling crossover. Our findings establish dual-gated TMD homobilayer as a versatile platform for on-chip, dynamically reconfigurable nonlinear light-matter physics.

## Introduction

Realizing strong, electrically tunable photon-photon interactions in solid-state systems is a long-standing goal for quantum photonics, underpinning the development of programmable simulators and logic gates (*1*, *2*). Exciton-polaritons—bosonic quasiparticles formed by the strong coupling between cavity photons and material excitons—offer a compelling platform for this pursuit (*3*). By inheriting the small effective mass of their photon component and the strong nonlinear interactions of their excitonic counterpart, polaritons can manifest macroscopic collective phenomena, including low-threshold lasing (*4*), polariton condensation (*5*–*8*), and superfluidity (*9*).

Monolayer transition metal dichalcogenides (TMDs) have recently emerged as appealing systems for studying exciton-polaritons, featuring tightly bound intralayer excitons with large oscillator strength (*10*–*13*). However, their practical utility is often hindered by weak, short-range interactions stemming from the small Bohr radius of intralayer excitons and their limited lifetimes (*14*–*16*). Various strategies, including the utilization of trions (*17*), polarons (*18*), biexcitons (*19*), Rydberg states (*20*), and the confinement potentials in traps (*21*) or moiré superlattices (*22*), have been employed to enhance nonlinearity. Yet, a critical bottleneck remains: active control over the polariton properties is typically limited once microcavity devices are fabricated (*23*). For the next generation of integrated polaritonic devices, tunable and reconfigurable nonlinear interactions are highly desirable. Although electrostatic doping has been widely used to modulate intralayer polaritons (*17*, *18*, *24*, *25*), integrating *in situ* electrical field control is essential to transcend from observing passive phenomena to engineering programmable photonic circuits, as previously demonstrated in III-V quantum well systems (*26–29*).

To overcome the weak, short-range interactions characteristic of intralayer excitons and provide an additional route to electrically control excitonic interactions, interlayer excitons (IEs) in bilayer TMDs offer a promising platform by separating electrons and holes into adjacent layers (*30*). This charge separation creates permanent out-of-plane electric dipoles, which exhibit long-range interactions (*31*, *32*), extended lifetimes (*33*), and high electrical tunability (*34*, *35*). Nevertheless, IEs in TMD heterobilayers usually suffer from weak oscillator strength, which is typically two orders of magnitude lower than that of intralayer excitons due to spatial electron-hole separation, precluding the realization of dipolar polaritons in the robust strong coupling regime (*36–38*).

A distinct solution to overcome this oscillator strength bottleneck is offered by the naturally stacked 2H-$MoS_2$ homobilayer, where two identical monolayers are rotated 180° relative to each other, retaining inversion symmetry (Fig.1A). In this system, the hole is delocalized across the layers mediated by coherent tunneling, resulting in the hybridization of the momentum-direct intralayer B exciton and momentum-indirect IEs (Fig.1B) (*39–41*). Consequently, unlike the heterobilayer counterparts, IEs in $MoS_2$ homobilayer exhibit both strong oscillator strength and electric-field tunability. Recent studies have demonstrated the giant Stark effect (*39–42*), electrically tunable ultrafast dynamics and interactions in bare IEs (*43*, *44*), and the formation of static dipolar polaritons (*45–47*). However, the dynamic electrical control of the fundamental parameters governing dipolar polaritons—such as the light-matter coupling strength and effective nonlinear coefficient—has remained largely unexplored in the two-dimensional semiconductor microcavities.

Here, we integrate a dual-gated bilayer $MoS_2$ device into a planar optical microcavity to realize an electrically reconfigurable dipolar-polariton platform. Building on previous demonstrations of dipolar polaritons, our device architecture enables continuous reconfiguration of the polariton dispersion and tuning of the light-matter coupling strength *via* the quantum-confined Stark effect. This control yields a seven-fold enhancement of the effective polariton nonlinear coefficient compared to the zero-field baseline. The enhancement results from the combined effects of the field-dependent effective excitonic nonlinearity and the excitonic Hopfield coefficient. Furthermore, by exploiting carrier density as an additional degree of freedom, we implement an all-electrical control that continuously drives the system across a strong-to-weak coupling crossover. This extensive control over dipolar polaritons provides a route toward electrically reconfigurable polaritonic devices and dynamically tunable nonlinear light-matter functionalities.

## Results

### Intrinsic excitonic landscape and electrical tunability of bilayer $MoS_2$.

In our experiment, a naturally stacked 2H bilayer $MoS_2$ is encapsulated within hexagonal boron nitride (hBN) flakes and integrated into a dual-gate structure, as schematically illustrated in Fig. 1C. Monolayer graphene serves as the top and bottom gate electrodes, specifically selected for high transparency to minimize optical absorption losses in subsequent cavity integration steps (*48*). The bilayer $MoS_2$ is electrically contacted to ground through a few-layer graphite electrode. This device architecture enables independent control over the out-of-plane electric field $E_z$ and the electrostatic doping density $n$ within the sample (see Supplementary Note 1) (*49*). Figure 1D shows an optical micrograph of a representative device. For subsequent cavity integration, this heterostructure serves as the active medium in a planar microcavity, composed of a 9-period distributed Bragg reflector (DBR) bottom mirror and a 50 nm silver top mirror (see Methods for fabrication details).

We first characterize the intrinsic excitonic landscape of the bilayer $MoS_2$ by performing reflectivity spectroscopy at 9 K in a half-cavity configuration, before transfer of the top mirror. At zero electric field ($E_z = 0$ V/nm, black curve in Fig. 1E), the spectrum exhibits three dominant resonances corresponding to the ground-state A exciton ($A_{1s}$) at 1.929 eV, the B exciton at 2.106 eV, and the IE at 1.996 eV. The IE retains a substantial oscillator strength, approximately 35% of the $A_{1s}$ resonance. Upon applying a vertical electric field of $E_z = 0.03$ V/nm (red curve in Fig. 1E), the single IE resonance splits into two distinct branches, denoted $IE_L$ and $IE_H$ for the lower- and higher-energy branches, respectively. The two hybrid IE branches carry effective out-of-plane dipole moments with opposite orientations (*50*). The field evolution of these resonances is mapped in Fig. 1F, where the two IE branches exhibit a pronounced X-shaped Stark splitting. In contrast, the intralayer $A_{1s}$ and B exciton remain largely insensitive to the field, confirming the dominant interlayer character of the IE state (*39*, *41*).

From the linear Stark shifts, we extract the effective electric dipole moments $p_{\text{eff}}$ to be around 0.40 $e$·nm for both $IE_L$ and $IE_H$ (see Supplementary Note 2). By benchmarking the measured effective dipole moment $p_{\text{eff}}$ against the theoretical limit of a pure interlayer state, $p_0 \approx 0.65$ $e$·nm (determined by the geometric layer separation $d_0 \approx 0.65$ nm) (*35*), we estimate the interlayer component weight $|C_{\text{IE}}|^2 = p_{\text{eff}}/p_0 \approx 0.61$. This result indicates that the observed state is a hybridized admixture of the pure IE and the intralayer B exciton, consistent with previous theoretical and experimental studies attributing this hybridization to coherent hole tunneling (*39–41*). The intralayer component therefore provides substantial oscillator strength while the IE component retains a sizeable out-of-plane dipole moment, providing the ingredients required for strong light-matter coupling. Additionally, we observe a weak resonance near 2.072 eV with a field-dependent shift that follows the trend of the ground-state IE resonance (Fig. 1F). We attribute this resonance to the first excited Rydberg state of the IE, denoted $IE_{2s}$ (see Supplementary Note 3) (*51*).

We next explore the independent knob of electrostatic doping by varying the carrier density $n$ at zero electric field (Fig. 1G). The system exhibits a pronounced asymmetry between electron and hole doping. Upon electron doping ($n > 0$), the neutral $A_{1s}$ resonance evolves into a blue-shifted and broadened repulsive polaron (RP) branch, which rapidly loses oscillator strength due to phase-space filling and exchange interactions with the Fermi sea (*18*, *52*). Concurrently, the spectral weight is transferred to a redshifted attractive polaron (AP) branch, which emerges at lower energies (*24*, *53*). The IE resonance is also strongly suppressed and eventually disappears at higher electron densities, consistent with the occupation of *K*-valley conduction-band states that directly induce phase-space filling and screening of the *K*-valley optical transitions (*54*, *55*).

In contrast, under hole doping ($n < 0$), both the $A_{1s}$ and IE resonances largely retain their substantial oscillator strength across the entire probed density range. We attribute this pronounced electron-hole asymmetry to the band structure of bilayer $MoS_2$, where the valence band maximum is located at the Γ point of the Brillouin zone, approximately 400 meV above the local maximum at the *K* point, as established by previous angle-resolved photoemission spectroscopy and theoretical studies (*56*, *57*). Within our experimental doping range ($|n| \leq 6 \times 10^{12}$ cm$^{-2}$), quantitative estimates indicate that the Fermi level shifts downward by only ~24 meV (see Supplementary Note 4), far less than the *Γ*-*K* valence band offset. The injected holes therefore remain predominantly in the *Γ* valley and are kinetically decoupled from the *K*-valley optical transitions. The resulting momentum mismatch suppresses Pauli blocking and exchange interactions, preserving the oscillator strength of IE states.

**Electrically tunable nonlinearity and ultrafast dynamics in bilayer $MoS_2$.**

Having established the steady-state electrical control of the excitonic resonances, we next employ ultrafast pump-probe spectroscopy to investigate the nonlinear response and dynamics of the hybrid IEs. A narrow-band pump pulse centered at 2.070 eV with a full width at half maximum (FWHM) of 50 meV is used to resonantly excite the B-exciton manifold, while a delayed broadband supercontinuum pulse probes the transient response of the bilayer $MoS_2$ (see Methods for details). Figure 2A and 2B display the pump-fluence-dependent transient reflectivity spectra of the $IE_L$ resonance at a fixed delay of 200 fs, where zero delay time is defined in Methods. At $E_z = 0$ V/nm (Fig. 2A), increasing pump fluence induces a small blueshift of $IE_L$ (up to ~1.4 meV), accompanied by noticeable peak broadening and bleaching. In contrast, at $E_z = 0.07$ V/nm (Fig. 2B), the resonance exhibits a pronounced density-dependent blueshift of up to ~5.2 meV.

To understand the origin of this field-dependent nonlinear response, we examine how the electric field modifies the hybridization character of the two Stark-split IE branches (Fig. 2C, D). For the $IE_L$ branch (Fig. 2C), the applied field increases the energy offset between the IE and the B exciton, thereby reducing their hybridization and increasing the interlayer character of the state (see Supplementary Note 5) (*58*). This reduced hybridization is accompanied by a decrease in oscillator strength (see Supplementary Note 6) and an increase in the effective dipole moment ($p_{\mathrm{eff}}$) from 0.40 $e$·nm to 0.48 $e$·nm. By contrast, for the $IE_H$ branch (Fig. 2D), the applied field brings the hybrid interlayer exciton closer to the bright intralayer B exciton, increasing its intralayer character.

We then quantify the nonlinear response of $IE_L$ by extracting the energy shift, $\Delta E_{\mathrm{L}}$, as a function of the exciton density, $n_{\mathrm{exc}}$, using Lorentzian lineshape analysis (Fig. 2E). The magnitude of blueshift exhibits a pronounced dependence on the applied field. We define an effective exciton nonlinear coefficient, $g_{\mathrm{exc}}^{\mathrm{eff}}$, from the low-density slope, $\Delta E_{\mathrm{L}} = g_{\mathrm{exc}}^{\mathrm{eff}} n_{\mathrm{exc}}$. Here, $g_{\mathrm{exc}}^{\mathrm{eff}}$ characterizes the measured energy shift of the IE resonance under the present finite-bandwidth excitation and subsequent relaxation conditions, rather than a bare microscopic two-body interaction constant. The extracted $g_{\mathrm{exc}}^{\mathrm{eff}}$ increases monotonically from 0.29 to 1.81 μeV·μm$^2$ as the field is increased (Fig. 2E inset; see Supplementary Note 7 for details). This corresponds to a continuous six-fold enhancement of the effective nonlinear coefficient over the investigated field range. The enhancement is consistent with the increased interlayer character and effective dipole moment of the $IE_L$ branch under vertical electric field. However, the associated increase in dipolar repulsion, estimated as $g_{\mathrm{dd}} \propto {p_{\mathrm{eff}}}^2 \propto |C_{\mathrm{IE}}|^4$, accounts for only a factor of ~1.4 (see Supplementary Note 8), indicating that the full six-fold modulation cannot be explained by the dipole enhancement alone.

The larger modulation of $\Delta E_{\mathrm{L}}$ can be understood by considering the population balance between the two oppositely oriented IE dipoles. Within a phenomenological Hartree-like description, the energy shift of the $IE_L$ can be expressed as $\Delta E_{\mathrm{L}} = g_{\mathrm{LL}} n_{\mathrm{L}} + g_{\mathrm{LH}} n_{\mathrm{H}}$, where $n_{\mathrm{L}}$ ($n_{\mathrm{H}}$) denotes the population density for the $IE_L$ ($IE_H$), and $g_{\mathrm{LL}}$ ($g_{\mathrm{LH}}$) is the intra- (inter-) species interaction coefficients. Because bandgap renormalization effects are strongly suppressed in bilayer $MoS_2$ compared with monolayer systems (*43*, *59*, *60*), we use a dipolar-interaction picture to interpret the dominant density-dependent contribution to the observed energy shift. The intraspecies interaction between aligned dipoles is repulsive, $g_{\mathrm{LL}} \propto {p_{\mathrm{L}}}^2 > 0$, whereas the interspecies interaction between oppositely oriented dipoles is attractive, $g_{\mathrm{LH}} \propto p_{\mathrm{L}} p_{\mathrm{H}} < 0$ (see Supplementary Note 8). At zero electric field, the two degenerate IE branches are expected to be approximately balanced in population ($n_{\mathrm{L}} \approx n_{\mathrm{H}}$), so the repulsive interaction within the same branch ($g_{\mathrm{LL}} n_{\mathrm{L}}$) is compensated by the attractive interaction between opposite dipoles ($g_{\mathrm{LH}} n_{\mathrm{H}}$), consistent with the small blueshift observed at $E_z = 0$ V/nm and with recent many-body calculations (*60*). The residual small blueshift may arise from the short-range exchange interactions or local symmetry breaking due to structural disorder (*45*, *46*).

At finite electric fields, the Stark splitting lifts the degeneracy between the two IE branches and energetically separates the $IE_L$ and $IE_H$ reservoirs. Following excitation of the B-exciton manifold, the photoexcited population relaxes into these Stark-split IE reservoirs. Thermodynamic analysis and steady-state photoluminescence measurements indicate preferential population of the lower-energy $IE_L$ branch at finite field (see Supplementary Note 9). This preferential population would reduce the compensation from the attractive $g_{\mathrm{LH}} n_{\mathrm{H}}$ term and thereby enhance the net repulsive shift of $IE_L$. Therefore, the observed field enhancement of the effective nonlinear coefficient of $IE_L$ is consistent with two concurrent contributions: the increase in the effective dipole moment of the $IE_L$ branch, and the field-induced population redistribution between the two IE branches. Within a simple capacitor-model estimate for aligned interlayer dipoles, the corresponding interaction scale is ~1.8 μeV·μm$^2$ (see Methods for details) (*61*) , providing an interaction scale comparable to the maximum effective nonlinear coefficient extracted here, $g_{\rm exc}^{\rm eff} = 1.81\ \mu\text{eV} \cdot \mu\text{m}^2$.

For comparison, the $IE_H$ branch exhibits a weaker and non-monotonic density-dependent energy shift, consistent with its opposite hybridization trend and increased intralayer character at finite field (see Supplementary Note 10). This behavior further supports the branch-dependent nature of the nonlinear response. A benchmark against representative monolayer and bilayer systems is given in Supplementary Note 11.

The applied electric field also modifies the ultrafast carrier dynamics. As shown in the short-delay inset of Fig. 2F, higher fields prolong the relaxation of the transient blueshift, consistent with an enhanced interlayer character and a longer-lived dipolar exciton population. Over longer timescales (t > 50 ps, main panel), the initial blueshift evolves into a delayed redshift, consistent with transient lattice heating and thermal bandgap shrinkage (see Supplementary Note 12) (*62*). Measurements on an independent dual-gated device show similar field-dependent interaction trends and temporal profiles, supporting the reproducibility of the effect beyond a single device (see Supplementary Note 13).

Electrically tunable ultrafast dynamics and nonlinear responses of bare IEs in bilayer $MoS_2$ have been reported previously (*43*). Here, we extend these studies by extracting the effective nonlinear coefficient of the hybrid $IE_L$ as a continuously field-dependent quantity, supporting a role of field-induced population imbalance in the enhanced response, and resolving the temporal evolution of the nonlinear signal over the measured 340 ps time window. These results establish the field-dependent effective excitonic nonlinear response that forms the basis for the polariton studies described below.

**Electrically reconfigurable strong coupling in a planar microcavity.**

We next integrate the dual-gated heterostructure into a planar microcavity to investigate the electrical control of strongly coupled light-matter states. The microcavity is formed by transferring a 50 nm Ag top mirror above a poly-methyl methacrylate (PMMA) spacer designed to position the bilayer $MoS_2$ close to the cavity-field antinode. At $k_\parallel/k_0 = 0$, the fundamental cavity mode is negatively detuned by ~ 47 meV from the hybrid IE resonance, where $k_\parallel$ and $k_0$ denote the in-plane and free-space wave vectors, respectively. The cavity exhibits a moderate quality factor because of optical losses associated with the metallic mirror and graphene contacts; a detailed characterization of the cavity is provided in Supplementary Note 14.

Angle-resolved reflectivity spectroscopy at $E_z = 0$ V/nm reveals four distinct polariton branches (Fig. 3A), consistent with the simultaneous coupling of a single cavity mode to the intralayer A exciton, hybrid IE, and intralayer B exciton (*45*, *46*). By fitting the dispersion with a four-coupled-oscillator model (see Supplementary Note 15), we extract the Rabi splittings, yielding $\Omega_{\mathrm{A}} = 56.2$ meV, $\Omega_{\mathrm{IE}} = 30.4$ meV, and $\Omega_{\mathrm{B}} = 53.6$ meV, respectively. Importantly, these values satisfy the strong coupling criterion $\Omega > \gamma_{\mathrm{cav}} + \gamma_{\mathrm{exc}}$, where $\gamma_{\mathrm{cav}}$ and $\gamma_{\mathrm{exc}}$ are the cavity and exciton linewidths, respectively (*10*, *63*). Together, the resolved polariton branches, coupled-oscillator fits, and linewidth comparison establish strong coupling of the cavity mode to all three excitonic resonances, including the optically active hybrid IE (see Supplementary Note 16).

Leveraging the Stark effect, we dynamically reconfigure the polariton dispersion. As the applied electric field splits the bare IE resonance into dipolar $\mathrm{IE_L}$ and $\mathrm{IE_H}$ branches, the polariton spectrum evolves from a four-branch to a five-branch structure (Fig. 3B). The full field-dependent evolution of the angle-resolved dispersion is presented in Supplementary Note 17. To visualize this electrical control, we integrate the reflectivity over a momentum window of $k_{\parallel}/k_0 = 0.31 \pm 0.03$, close to the relevant cavity-exciton crossing region, and plot the resulting spectra as a function of electric field in Fig. 3C. The polariton branches associated with the two IE resonances evolve continuously and symmetrically with field, reflecting the underlying X-shaped Stark effect. A five-coupled-oscillator model captures the full field-dependent polariton evolution. The extracted energies of the $\mathrm{IE_L}$ and $\mathrm{IE_H}$ agree well with those measured independently in the half-cavity configuration (see Supplementary Note 18), supporting the assignment of the observed polariton branches and the reliability of the fitted exciton energies. Notably, our analysis reveals that the light-matter coupling strength $V$ itself is highly tunable *via* the electric field. As extracted from the model and plotted in Fig. 3D, the coupling strength $V_{\mathrm{IE_H}}$ for $\mathrm{IE_H}$ branches nearly doubles as the electric field sweeps from 0 to 0.08 V/nm. This enhancement stems from the field-driven hybridization. As the field tunes the $\mathrm{IE_H}$ closer to the bright intralayer B exciton, it acquires large oscillator strength through hybridization, as shown in Fig. 1F and Supplementary Fig. S4. In contrast, the $V_{\mathrm{IE_L}}$ experiences a slight decrease to $\sim$ 12 meV under finite electric field. The corresponding full Rabi splitting, $\Omega_{\mathrm{IE_L}} = 2V_{\mathrm{IE_L}} \approx 24$ meV, remains above the combined linewidth threshold within the measured field range (see Supplementary Note 16). Thus, the vertical electric field reconfigures the polariton Hamiltonian through the simultaneous tuning of the exciton energies, exciton-cavity detuning, and light-matter coupling strength.

Additionally, we demonstrate that strong coupling can be selectively quenched *via* electrostatic doping (see Supplementary Note 19 for the full spectral evolution). Figure 3E maps reflectivity at $k_{\parallel}/k_0 = 0.35 \pm 0.03$ as a function of doping density. Consistent with the half-cavity device measurements in Fig. 1G, electron doping progressively suppresses the IE oscillator strength through Pauli blocking and screening (*64*). Correspondingly, the extracted coupling strength $V_{\mathrm{IE}}$ decreases continuously with electron density, causing the IE-associated upper and lower polariton branches to approach each other and eventually become unresolved as separate resonances. This continuous reduction drives the system across the strong-coupling threshold into the weak-coupling regime (see Supplementary Note 20 for more details). The corresponding crossover is not observed over the measured hole-doping range, consistent with the substantially weaker suppression of the IE oscillator strength under hole doping. The resulting polarity dependence provides an additional electrical handle for controlling the IE-cavity coupling regime within the same device and offers a route toward active polariton gating and routing (*64*, *65*).

**Electrical tuning of the nonlinearity of the dipolar polaritons**

Finally, we use angle-resolved ultrafast pump-probe spectroscopy to investigate the nonlinear response of the electrically reconfigurable dipolar polaritons (*66*). We employ the same pump-probe configuration as in the bare exciton measurements, with the pump spectrally addressing the B-exciton manifold while non-resonantly populating the IE-associated polariton branches, and a delayed broadband supercontinuum pulse probing the transient reflectivity (see Methods and Supplementary Note 21). Figures 4A and 4B present the pump-density-dependent transient reflectivity spectra of the $\mathrm{IE_L}$-associated polariton branches at $E_z = 0$ V/nm and at a finite field $E_z = 0.04$ V/nm, respectively. In both regimes, increasing pump fluences induces a clear energy blueshift of the polariton resonances—a signature of repulsive polariton-polariton interactions enhanced by the dipolar character of the excitonic component.

To quantify the electrical control over the polariton nonlinearity, we extract the energy shift of the lower polariton branch (LPB) associated with the $\mathrm{IE_L}$ at near zero-detuning ($k_{\parallel}/k_0 = 0.31 \pm 0.03$) as a function of polariton density (Fig. 4C). The data is presented on a logarithmic density scale to visualize the broad dynamic range, while the linear-scale inset highlights the steep rise in the low-density regime. This rapid initial blueshift implies a large interaction strength before the onset of saturation. At higher densities, the shift rapidly saturates. We attribute this to the phase space filling effect, which reduces the Rabi splitting and counteracts the interaction-induced blueshift, consistent with previous reports on dipolar polaritons in bilayer $MoS_2$ (*45*). To characterize this sublinear response, we define a density-dependent effective polariton nonlinear coefficient, $g_{\mathrm{LP}} = \Delta E / n_{\mathrm{LP}}$, which parameterizes the measured polariton energy shift under the present excitation and relaxation conditions rather than a bare microscopic polariton-polariton interaction constant. Using this metric, we find that $g_{\mathrm{LP}}$ exhibits a remarkable dependence on the electric field. As plotted in Fig. 4D, $g_{\mathrm{LP}}$ increases from 0.09 to 0.60 $\mu eV \cdot \mu m^2$ in the low-density limit with the application of the electric field, representing a nearly seven-fold enhancement over the applied range. Given the rapid saturation due to phase-space filling effect (Fig. 4C, inset), the intrinsic nonlinearity in the dilute limit is expected to be larger. To contextualize the magnitude of the polariton nonlinearity in our device, we quantitatively benchmarked our extracted interaction strength against other representative polaritonic platforms (see Supplementary Note 22 and Table S2).

This macroscopic enhancement surpasses the scaling observed for bare excitons alone (compare Fig. 4D with Fig.2C). We interpret this amplification in terms of a synergistic dual mechanism. Firstly, the electric field enhances effective excitonic nonlinear response $g_{\mathrm{exc}}^{\mathrm{eff}}$, consistent with the field-induced population imbalance where excitons preferentially occupy the strongly dipolar lower branch ($n_{\mathrm{L}} \gg n_{\mathrm{H}}$) and the increased effective dipole moment, as discussed in the previous section. Secondly, and unique to the cavity system, the electric field dynamically optimizes the polariton composition. The Stark-induced redshift of the exciton energy reduces the exciton-cavity detuning, thereby increasing the excitonic Hopfield coefficient ($|X|^2$) (see Supplementary Note 23). Crucially, while the underlying microscopic scattering in such cavities may be complex (*67*), the effective polariton-polariton interaction reliably scales with the fourth power of this excitonic fraction ($g_{\mathrm{LP}} \approx |X|^4 g_{\mathrm{exc}}$). Therefore, the field-driven increase in $|X|^2$ acts as a powerful lever to amplify the nonlinear interaction. Analysis at a non-zero detuning further validates this scaling, revealing a reduced blueshift consistent with the lower Hopfield fraction (see Supplementary Note 24).

Consequently, the seven-fold enhancement of $g_{\mathrm{LP}}$ is not merely a reproduction of the excitonic behavior, but the combined result of simultaneously boosting the microscopic repulsion and maximizing the macroscopic excitonic participation. Furthermore, simultaneous tracking of multiple polariton branches supports that the observed nonlinearity near zero detuning cannot be explained by Rabi-saturation effects alone and includes a substantial contribution from electrically enhanced dipolar repulsion (see Supplementary Note 25). By enabling a seven-fold modulation of the nonlinearity within a single device, this work complements static moiré and monolayer systems by adding in situ electrical reconfigurability, opening avenues for actively tunable polaritonic devices where interaction parameters can be dynamically reconfigured on demand.

## Discussion

In summary, we have realized an electrically reconfigurable dipolar-polariton platform by integrating a dual-gated bilayer $MoS_2$ heterostructure into a planar microcavity. We successfully bridge the gap between the highly tunable nonlinearity of interlayer states and the robust optical coupling of intralayer excitons. The central advance of this work lies in the device integration that enables simultaneous in situ electrical modulation of the light-matter coupling strength and the effective nonlinear response within a single device. The vertical electric field tunes both the effective excitonic interaction and the excitonic Hopfield fraction of the polariton, resulting in a nearly seven-fold modulation in the low-density effective polariton nonlinear coefficient over the measured field range. While this microscopic picture is consistent with the observed trends, the present measurements do not allow a unique quantitative decomposition of the enhancement into its individual contributions. Independently, electrostatic doping provides another control axis by progressively reducing the IE oscillator strength through Pauli blocking and screening, thereby continuously driving a strong-to-weak coupling crossover. The current blueshift-to-linewidth figure of merit remains limited by the moderate polariton linewidth. Further improvements in cavity quality and reductions in excitonic and device-related broadening would enhance the spectral visibility of the nonlinear energy shifts. Within these limitations, the dual-gated architecture enables in situ electrical control of the dipolar-polariton dispersion, effective nonlinear response, and light-matter coupling regime.

Our work positions dipolar polaritons based on TMD homobilayers as promising candidates for future studies of gate-reconfigurable nonlinear polaritonic devices (*68*, *69*). The demonstrated electrical control over dispersion, coupling strength, and effective nonlinear response may provide a route toward integrated polaritonic circuits and active sensing architectures (*28*). We also note that a related study investigating polariton-mediated binding in dipolar systems (*70*) appeared on the arXiv repository shortly after our initial report was posted, highlighting the growing interest in exploring tunable dipolar interactions within TMD homobilayer platforms.

## Materials and Methods

### Device and microcavity fabrication.

The bottom distributed Bragg reflector (DBR), consisting of 18 alternating layers of titanium dioxide ($TiO_2$) and silicon dioxide ($SiO_2$) layers, was deposited on a silicon substrate by electron-beam evaporation. Standard lithography was subsequently used to pattern Cr/Au contacts on the DBR surface to define the gate electrodes.

The heterostructure was assembled using mechanically exfoliated 2H-stacked bilayer $MoS_2$, monolayer graphene, and hBN flakes. Monolayer graphene was chosen for the top and bottom gate electrodes to minimize optical absorption, thereby mitigating the degradation of the cavity quality factor. The 2H-stacked bilayer $MoS_2$ flake, identified and confirmed by its characteristic reflectance and Raman spectra, was encapsulated between hBN flakes with nearly identical thickness (~12 nm) serving as the gate dielectrics. The stack was then assembled in a graphene/hBN/bilayer $MoS_2$/hBN/graphene sequence using a polycarbonate (PC)-based dry transfer technique, and subsequently transferred onto the DBR substrate, with the graphene gates precisely aligned to the pre-fabricated contacts.

To complete the $\lambda/2$ microcavity, a poly-methyl methacrylate (PMMA) spacer was spin-coated over the heterostructure. Finally, a 50 nm silver film, serving as the top mirror, was transferred onto the PMMA using a polydimethylsiloxane (PDMS, PF-X4, Gel-Pack) stamp-assisted method.

### Optical measurements.

**Experimental setup.** All optical measurements were performed at 9 K using a home-built, angle-resolved confocal spectral setup based on a microscope (Olympus). The sample was mounted in a closed-cycle cryostat (CIA). A 50× objective with a numerical aperture (NA) of 0.7 was used for both sample excitation and signal collection. The signal was dispersed by a Horiba iHR550 spectrometer, equipped with interchangeable 300 and 600 grooves/mm gratings, and detected by a liquid-nitrogen-cooled charge-coupled device (CCD, Symphony II). For all experiments, gate voltages were supplied by a source meter (Keithley 2614B). The system can be configured to image either the real plane or the Fourier plane of the objective onto the spectrometer entrance slit to acquire real-space or angle-resolved (k-space) spectra, respectively.

**Steady-state reflectivity.** The device was illuminated with a broadband halogen lamp (Thorlabs OSL2IR). Reflectivity contrast spectra were obtained by normalizing the signal from the device area to a reference spectrum from the substrate.

**Transient reflectivity spectroscopy.** The ultrafast optical response was investigated using a pump-probe configuration integrated into the confocal system. The light source was a regeneratively amplified Ti: sapphire laser system (Coherent, Astrella) delivering 100 fs pulses at 800 nm with a repetition rate of 5 kHz. The beam was split into pump and probe paths. The pump beam was converted to the desired wavelength using an optical parametric amplifier (OPA, Light Conversion) and spectrally shaped using a bandpass filter (FWHM ~ 50 meV) to ensure selective excitation. The probe beam was a broadband supercontinuum light (480-820 nm) generated in a sapphire crystal, with the time delay controlled by a motorized mechanical delay stage. The pump and probe beams were collinearly focused and spatially overlapped on the sample surface, yielding spot diameters of approximately 3 μm and 2 μm, respectively. Transient reflectivity spectra were recorded as a function of the pump-probe delay, with time zero defined by the temporal overlap of the pulses. We extract the transient signals at a fixed pump-probe delay of 200 fs. Because the pump does not directly excite the analyzed exciton and polariton branches, the measured energy shifts characterize an effective nonlinear response under indirect population and subsequent relaxation of the photoexcited exciton–carrier reservoir, rather than an instantaneous microscopic interaction. The choice of the 200-fs delay and the roles of carrier thermalization and competing many-body relaxation channels are discussed in Supplementary Note 26. For angle-resolved pump-probe spectroscopy, the back focal plane (Fourier plane) of the objective was imaged onto the spectrometer slit, enabling the mapping of transient carrier dynamics in energy-momentum space.

**Estimate of the exciton density.**

For the fluence-dependent measurements in the half-cavity configuration presented in Fig. 2E, we estimate the exciton density using the relation:

$$n_{\text{exc}} = \frac{P\alpha}{RA_{\text{laser}}E_{\text{pump}}} \tag{1}$$

where $P$ is the pump average power, $R$ is the repetition rate of the laser (5 kHz), $E_{\text{pump}}$ is the pump energy and $A_{\text{laser}} = \pi D^2/4$ is the size of the focused pump beam on the sample with a diameter $D$. The absorption coefficient $\alpha$ of bilayer $MoS_2$ was determined by convolving the reflectivity spectra and the laser profile. We assume all of the absorbed photons form excitons. We note that these fluences correspond to hybrid exciton densities (<

$10^{11}$ cm$^{-2}$) well below the exciton Mott transition ($\sim 10^{12}$ cm$^{-2}$).

**Estimate of polariton density**

To estimate the polariton density $n_{\mathrm{pol}}$ presented in Fig. 4, we calculated the injected photon density based on the pump parameters and accounted for the coupling efficiency into the polariton modes. The injected polariton density during the pulse is estimated by:

$$n_{\mathrm{pol}} = \frac{P\alpha}{CRA_{\mathrm{laser}}E_{\mathrm{pump}}} \qquad (2)$$

Where $C$ is the photonic fraction of the polariton states, this estimation allows us to correlate the observed energy blueshifts directly with the density of quasiparticles in the microcavity. It is important to note that our calculation assumes a unitary conversion efficiency from the absorbed pump photons (resonant with the B exciton) to the lower polariton reservoir. In practice, non-radiative scattering channels and relaxation bottlenecks likely result in a conversion efficiency below 100%. Consequently, our estimated polariton density represents an upper bound. Since the interaction strength is derived as $g_{\mathrm{LP}} = \Delta E / n_{\mathrm{pol}}$ , any overestimation of the density implies that our reported $g_{\mathrm{LP}}$ values represent conservative lower bounds of the intrinsic nonlinearity, and the actual interaction strength may be even larger.

**Estimate of the dipolar interaction strength.**

Based on a parallel-plate model (*61*), the dipole-dipole interaction strength is given by $g_{\mathrm{dd}} = e \cdot p_{\mathrm{eff}} / \varepsilon_0 \varepsilon_{\mathrm{eff}}$, where $p_{\mathrm{eff}}$ is the dipole moment of IE, which can be acquired from the Stark shift of IE and estimated to be 0.40-0.48 $e$·nm (see Supplementary Note 2). Theoretically, the effective permittivity is expected to lie close to the geometric mean of the heterostructure layers ($\varepsilon_{\mathrm{eff}} \approx \sqrt{\varepsilon_{\mathrm{TMD}} \cdot \varepsilon_{\mathrm{hbn}}} = 4.94$), thereby leading to an estimated dipolar interaction strength of $g_{\mathrm{dd}} \approx 1.8$ μeV·μm$^2$. This value aligns well with nonlinear estimations reported for other hybridized exciton systems, reinforcing the physical validity of our extraction (See Supplementary Note 11).

**Acknowledgments**

We extend our sincere appreciation to Professor Long Zhang for the valuable discussions on this research.

**Funding:** Q. X. gratefully acknowledges funding support from the National Key Research and Development Program of China (Grant No. 2022YFA1204700), the National Natural Science Foundation of China (Grant No. W2541001, 12434011), and the New Cornerstone Science Foundation (NCI202508). K.W. and T.T. acknowledge support from the JSPS KAKENHI (Grant Numbers 21H05233 and 23H02052), the CREST (JPMJCR24A5), JST and World Premier International Research Center Initiative (WPI), MEXT, Japan.

**Author contributions:** Q.X. supervised the whole project. B.X. and Q.X. conceived the idea of the research. B.X. fabricated the devices with the help of Z.Sh., G.W. and Y.L. K.W. and T.T. provided the hBN crystals. Y.W. and B.X. established the optical paths. B.X. performed the steady-state and transient spectral measurements with the help of Y.W. B.X., G.W., and Y.W. analyzed the data with guidance from Q.X., Z.Su., and T.C.H.L. H.W. and H.L. provided software support. B.X. and Y.W. co-wrote the manuscript with input from all authors.

**Competing interests:** The authors declare no competing interests.

**Data, Code, and Materials Availability:** All data and code needed to evaluate and reproduce the results in the paper are present in the paper and/or the Supplementary Materials. This study did not generate new materials.

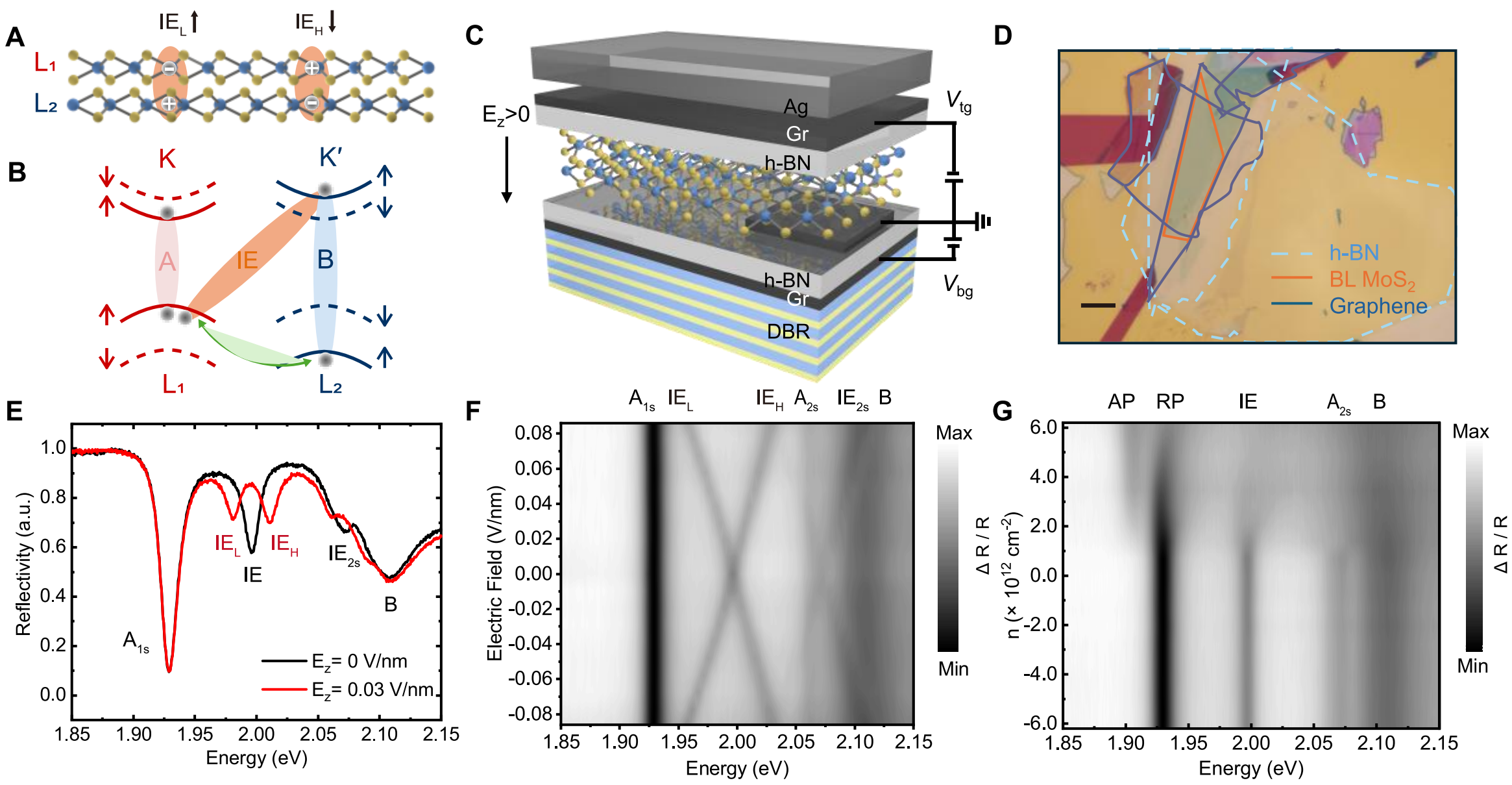

**Fig. 1. Dual-gated bilayer $MoS_2$ device and electrically tunable excitonic optical response at 9 K.** (**A**) Atomic structure and spatial charge distribution of interlayer excitons (IEs) in bilayer $MoS_2$. The top and bottom layers are denoted $L_1$ and $L_2$, respectively. (**B**) Electronic band alignment of 2H-stacked bilayer $MoS_2$ at the K point under zero electric field, showing the optical transitions associated with the intralayer A and B excitons and the hybrid IE. Hole tunneling between layers (green arrow) mediates hybridization between the IE and the B exciton. Spin states are indicated by arrows. (**C**) Schematic of the hBN-encapsulated bilayer $MoS_2$ device with graphene top and bottom gates, a bottom distributed Bragg reflector (DBR), and an Ag top mirror. (**D**) Optical micrograph of the device before transfer of the Ag top mirror. The bilayer $MoS_2$ (orange outline), hBN (blue outline), and graphene gates (gray outline) are indicated. Scale bar, 20 μm. (**E**) Reflectance-contrast spectra at $E_z$= 0 V/nm and $E_z$= 0.03 V/nm (black and red curves, respectively). At zero field, the intralayer A and B exciton and IE resonances are observed; at $E_z$= 0.03 V/nm, the IE resonance undergoes Stark splitting into the lower- and higher-energy branches, $IE_L$ and $IE_H$. (**F**) False-color map of the reflectance contrast as a function of applied electric field, showing the Stark splitting of the IE resonance, whereas the intralayer A and B excitons remain comparatively insensitive to the field. (**G**) Doping-dependent reflectance spectra at zero electric field. Under electron doping ($n > 0$), the A-exciton resonance evolves into attractive-polaron (AP) and repulsive-polaron (RP) branches, while the IE resonance is quenched. Under hole doping ($n < 0$), both the A-exciton and IE resonances retain substantial oscillator strength.

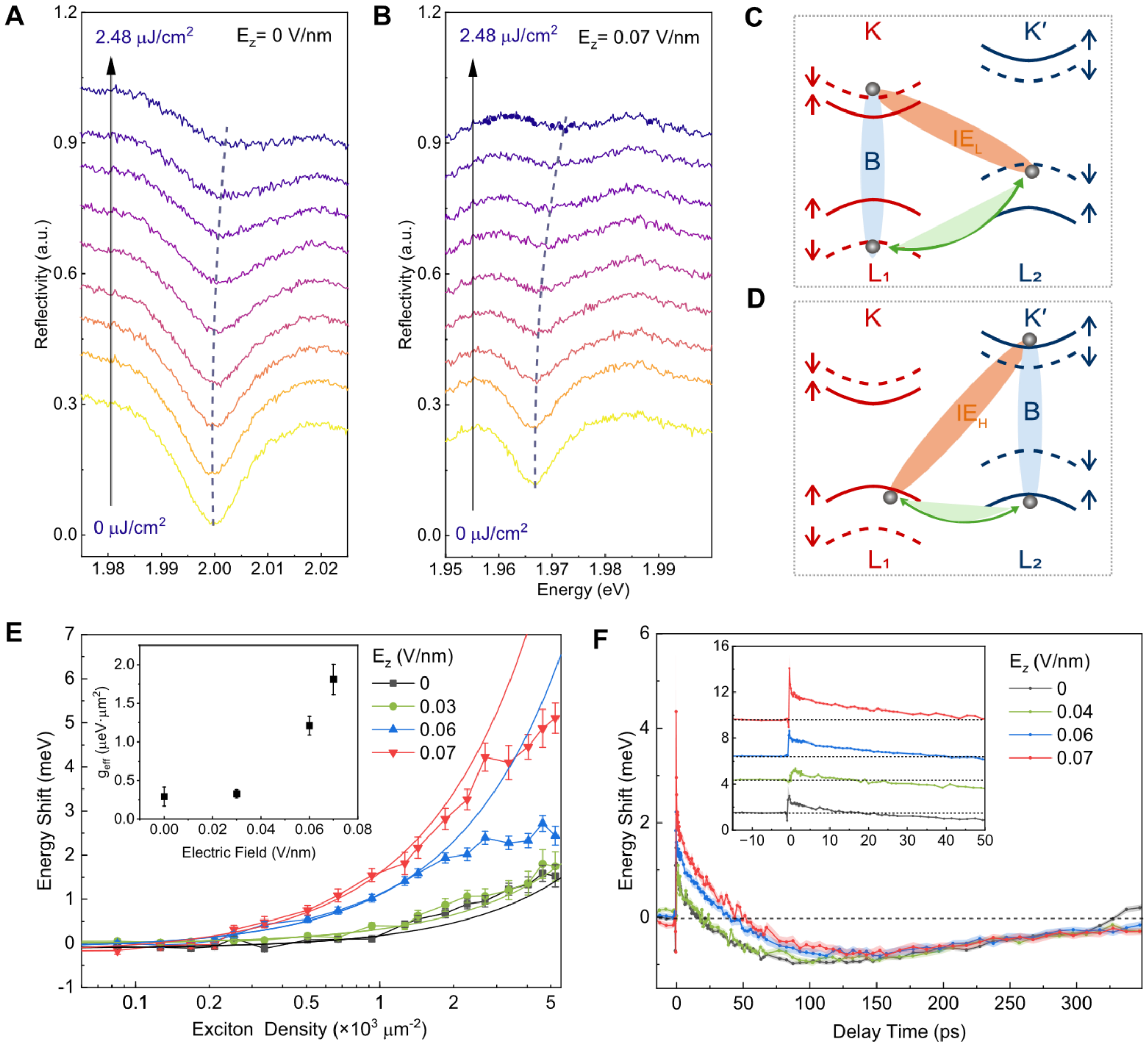

**Fig. 2. Electrically tunable exciton nonlinearity and ultrafast dynamics in bilayer $MoS_2$.** (**A and B**) Pump-fluence-dependent transient reflectance spectra of the IE (**A**) and $IE_L$ (**B**) resonances at a fixed delay of 200 fs under $E_z$ = 0 V/nm and $E_z$ = 0.07 V/nm, respectively. Dashed lines trace the resonance blueshift with increasing pump fluence. (**C and D**) Schematic illustration of the field-dependent hybridization mechanism. The applied electric field modifies the energy detuning between the interlayer-exciton and intralayer B-exciton basis states, thereby changing the interlayer character of the $IE_L$ branch in (C) and the $IE_H$ branch in (D), and consequently their nonlinear responses. (**E**) Extracted energy blueshift of the $IE_L$ resonance as a function of exciton density at different electric fields. Inset: Extracted effective exciton nonlinear coefficient as a function of $E_z$. (**F**) Temporal dynamics of the $IE_L$ energy shift at different electric fields over an extended 340-ps time window under the same pump power. Inset: Early-time dynamics over the first 50 ps, plotted with vertical offsets to highlight the field-dependent prolongation of the transient blueshift relaxation.

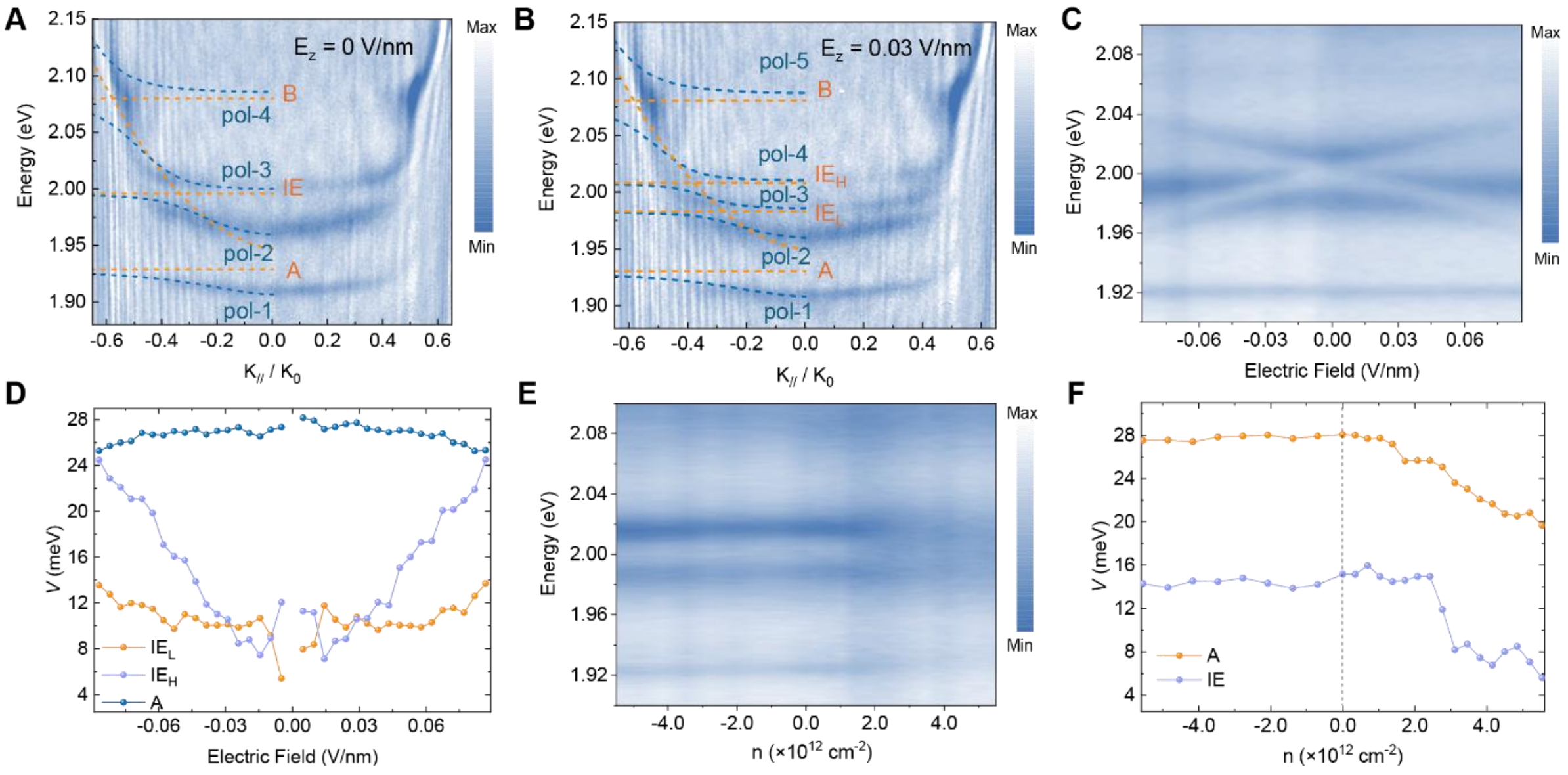

**Fig. 3. Electrically reconfigurable strong coupling and polariton dispersion in a dual-gated $MoS_2$ microcavity.** (**A** and **B**) Angle-resolved reflectivity spectra at (A) $E_z$= 0 V/nm and (B) $E_z$= 0.03 V/nm, respectively. Dashed lines are guides to the eye for the polariton eigenvalues (orange) and cavity modes (blue) obtained by fitting with a coupled-oscillator model. (**C**) False-color map of reflectivity spectra extracted at $k_\parallel/k_0$ = 0.31 as a function of applied electric field. (**D**) Extracted exciton–photon coupling strengths for the A exciton (blue), $IE_L$ branch (orange), and $IE_H$ branch (purple) as a function of electric field, obtained from the coupled-oscillator model. (**E**) False-color map of reflectivity spectra extracted at $k_\parallel/k_0$ = 0.35 as a function of doping density. (**F**) Extracted coupling strengths of the IE and A-exciton resonances as a function of doping density. Electron doping progressively reduces the IE oscillator strength and drives a strong-to-weak-coupling crossover.

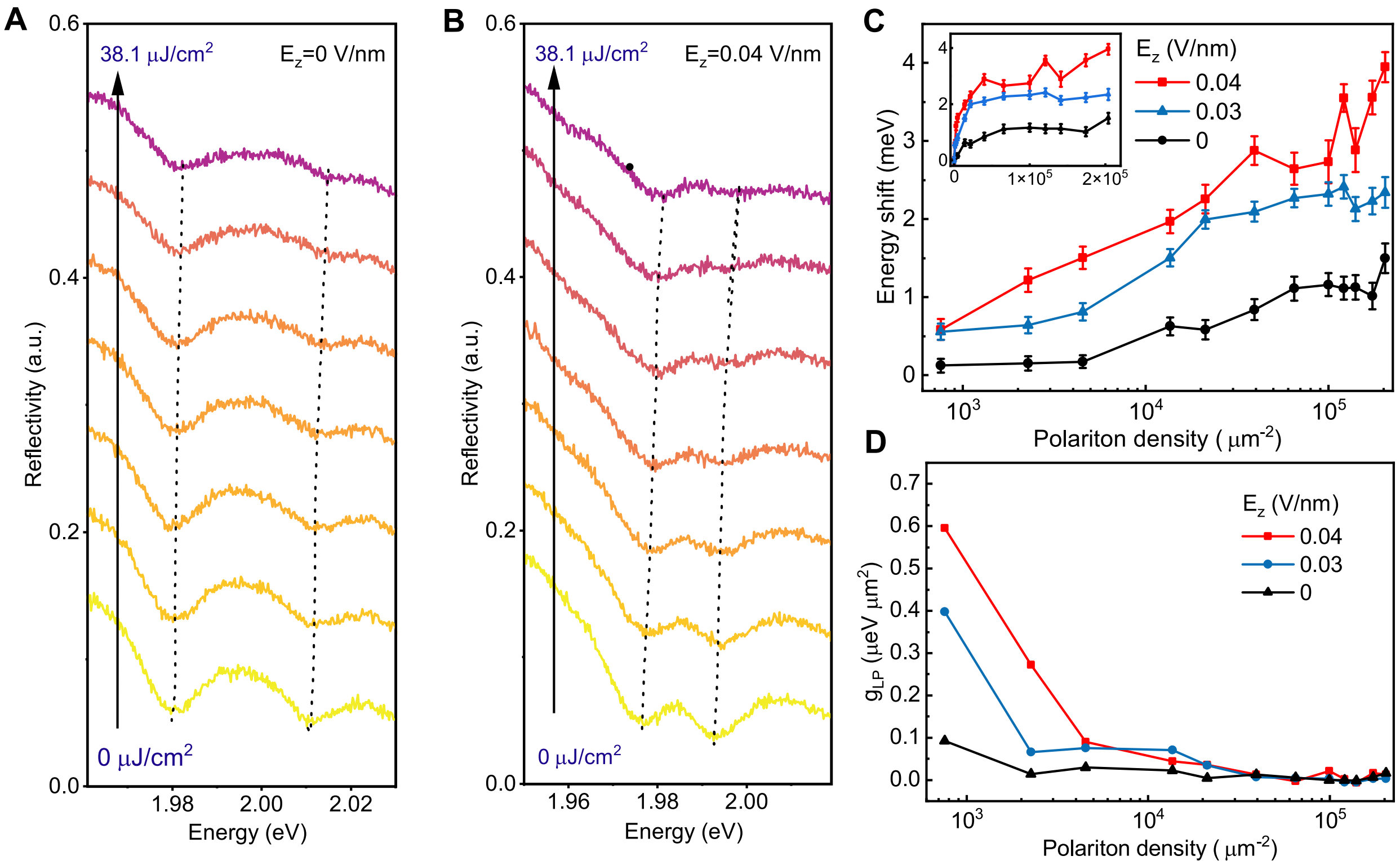

**Fig. 4. Electrical tuning of polariton nonlinearities.** (**A** and **B**) Resonant reflectivity spectra of the lower and upper polariton branches (LPB and UPB) of $IE_L$ at zero detuning for (A) $E_z$= 0 V/nm and (B) $E_z$= 0.04 V/nm, as a function of increasing pump-induced polariton density. Dashed lines are guides to the eye, highlighting the density-dependent blueshift. (**C**) Extracted energy shift of the $IE_L$-associated LPB at the zero-detuning wavevector as a function of polariton density for different applied electric fields. The main panel uses a logarithmic x-axis to illustrate the nonlinear behavior over a wide density range. The inset displays the data on a linear scale to clearly visualize the steep initial rise of the effective nonlinear response. Error bars represent the uncertainty in the extracted energy shifts from Lorentzian fits. (**D**) Effective polariton nonlinear coefficient $g_{LP}$ as a function of polariton density. The low-density interaction strength at $E_z$= 0.04 V/nm is almost sevenfold larger than that at $E_z$= 0 V/nm.